\documentclass{jfm}

\graphicspath{{figures/}}
\usepackage{graphicx}
\usepackage{natbib}
\usepackage{hyperref}
\usepackage{newtxtext,newtxmath}

\newcommand{\vect}[1]{\boldsymbol{#1}}
\newcommand{\DR}{\mathcal R}
\newcommand{\Pin}{\mathcal P_{in}}
\newcommand{\NS}{\mathcal S}

\title{Spatial discretization effects in spanwise forcing for turbulent drag reduction}
\author[E. Gallorini, M. Quadrio]
{Emanuele Gallorini and Maurizio Quadrio}
\affiliation{Dipartimento di Scienze e Tecnologie Aerospaziali, Politecnico di Milano, via La Masa 34, 20156 Milano, Italy}

\begin{document}

\maketitle
	
\begin{abstract}
Wall-based spanwise forcing has been experimentally used with success by Auteri et al. ({\em Phys. Fluids} vol. 22, 2010, 115103) to obtain large reductions of turbulent skin-friction drag and considerable energy savings in a pipe flow. 
The spatial distribution of the azimuthal wall velocity used in the experiment was not continuous, but piecewise-constant. 
The present study is a numerical replica of the experiment, based on a set of direct numerical simulations (DNS); its goal is the identification of the effects of spatially discrete forcing, as opposed to the idealized sinusoidal forcing considered in the majority of numerical studies.

Regardless of the discretization, with DNS the maximum drag reduction is found to be larger: the flow easily reaches complete relaminarization, whereas the experiment was capped at 33\% drag reduction.
However, the key result stems from the observation that, for the piecewise-constant forcing, the apparent irregularities of the experimental data appear in the simulation data too. They derive from the rich harmonic content of the discontinuous travelling wave, which alters the drag reduction of the sinusoidal forcing.
A detailed understanding of the contribution of each harmonic reveals that, whenever e.g. technological limitations constrain one to work far from the optimal forcing parameters, a discrete forcing may perform very differently from the corresponding ideal sinusoid, and in principle can outperform it. 
However, care should be exercised in comparison, as discrete and continuous forcing have different energy requirements. 
\end{abstract}

\section{Introduction}
\label{sec:introduction}

Turbulent flow control for drag reduction is an active discipline, pursuing a technological goal of steadily increasing economic and environmental importance.
A subset of flow control studies aims at reducing the drag of wall-bounded turbulent flows. This is a particularly difficult challenge: once geometry is simplified to consider a parallel flow, drag is only due to friction, and high levels of turbulent friction stem from the interaction between turbulence and the wall. 
Altering this interaction to improve the overall energetic efficiency, by following either active or passive approaches, is as fundamentally appealing as practically difficult.

An active, open-loop strategy to reduce friction drag which gained popularity in recent years is made by spanwise forcing \cite[a thorough review has been recently provided by ][]{ricco-skote-leschziner-2021}, notably in its spatially distributed version made by the streamwise-travelling waves (StTW), where the spanwise velocity component $W$ at the wall is prescribed according to: 
\begin{equation}
W(x,t) = A \sin \left( \kappa_x x - \omega t \right)
\label{eq:StTW}
\end{equation}
in which $x,t$ are the streamwise coordinate and time, and $\kappa_x$ and $\omega$ are the streamwise wavenumber and frequency of the oscillating wave, which is thus characterized by a phase speed $c = \omega / \kappa_x$. 
StTW, introduced by \cite{quadrio-ricco-viotti-2009}, are interesting owing mainly to their good energetic performance, which is preserved at high Reynolds numbers \citep{gatti-quadrio-2016}. 
StTW are effective even in compressible and supersonic flows (\cite{ruby-foysi-2022, gattere-etal-2024}), and have been demonstrated to affect favorably the aerodynamic drag of a three-dimensional body \citep{banchetti-luchini-quadrio-2020}, up to the point that an actuation over a limited portion of the wing of an airplane in transonic flight reduces the {\em total} drag of the aircraft by nearly 10\% at a negligible energy cost \citep{quadrio-etal-2022}. 

\begin{figure}
\centering
\includegraphics[width=\textwidth]{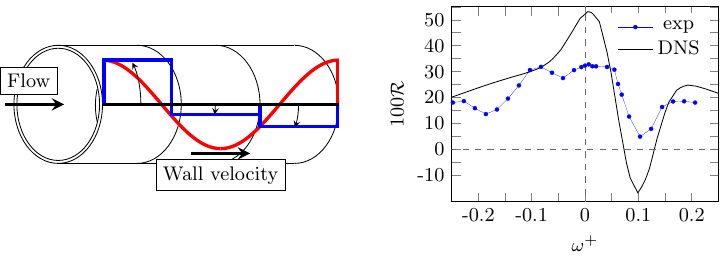}
\caption{Left: sketch of StTW actuation (continuous wave, red line) and DTW actuation with $s=3$ (discrete wave, blue line), as applied to the cylindrical pipe geometry by ABBCQ. Right: percentage drag reduction $\DR$ experimentally measured for DTW by ABBCQ in the pipe flow (symbols), compared at the same value of $\kappa^+$ to available DNS information for StTW from the DNS study of \cite{quadrio-ricco-viotti-2009} in the planar geometry (continuous black line). Besides the different geometry, the numerical study has a slightly smaller forcing amplitude and a slightly larger Reynolds number.} 
\label{fig:DTW}
\end{figure}

The major obstacle to the deployment of StTW (and of spanwise forcing in general) in practical applications is the lack of suitable actuators. 
Very few laboratory implementations of StTW are available, and typically the idealized sinusoidal waveform of the wall velocity cannot be achieved. 
\cite{bird-santer-morrison-2018} describe a planar actuator for StTW, formed by a tensioned membrane skin mounted on a kagome lattice; they discuss why the measured drag reduction turns out to be less than expected, and attribute it to the out-of-plane velocity component created by the actual forcing. 
\cite{benard-etal-2021} report preliminary results on the implementation of steady longitudinal waves of spanwise forcing via plasma DBD actuators, based on electrodes designed with a suitable shape, which are affected by the non-ideal response of the actuator. 
For a long time, the sole available laboratory experiment with StTW (and the most successful one) was the water pipe flow experiment carried out by \cite{auteri-etal-2010} (referred to as ABBCQ in this paper), who set up a low-Reynolds turbulent pipe flow modified by StTW and reported up to 33\% of drag reduction. 
StTW were implemented with an original device, recently replicated by \cite{marusic-etal-2021} in the plane geometry, in which the circular pipe is subdivided into axial slabs that independently oscillate in the azimuthal direction.
Different rotational speeds of nearby slabs provided the desired streamwise variation of transverse velocity, realizing a discrete travelling wave (DTW). A sketch of the DTW concept is provided in figure \ref{fig:DTW} (left). The experimental setup of ABBCQ consisted of 10 sets of 6 independently movable segments and allowed to test waves made by a variable number $s=2,3,6$ of segments discretizing each sinusoid. As emphasized in figure \ref{fig:DTW}, the actual streamwise distribution of the forcing spanwise velocity with DTW is far from the sinusoidal one (red line) and is instead well approximated by a staircase function (blue line).

The design of the ABBCQ pipe flow experiment was guided by the original numerical study by \cite{quadrio-ricco-viotti-2009}, carried out for the plane channel flow; however, the two flows are not identical. 
The drag reduction datasets, while in broad agreement, do present a few differences which have not been examined in depth so far.
Figure \ref{fig:DTW} (right) compares the experimental drag reduction data measured by ABBCQ for $s=3$ to the corresponding subset of DNS data, at the same wavelength of the travelling wave, with a similar forcing amplitude and Reynolds number.
The largest drag reduction rate measured in the experiment is 33\% instead of more than 50\% of the DNS. No drag increase is found in the experiment, and only a drop of drag reduction to nearly zero level is observed, albeit at the very same frequency $\omega^+ \approx 0.1$ where the DNS study predicts drag reduction to become negative. 
Moreover, the experimental data present a rather irregular, wiggling dependence upon $\omega^+$ compared to the smooth evolution of the DNS plane channel: especially at negative frequencies, some of the experimentally measured drag reduction rates are comparable and even marginally larger than the simulations, but other measurement points lie well below the numerical curve.

ABBCQ discuss some possible reasons behind these differences. The geometry (plane channel vs circular pipe) certainly does play a role; the value of the Reynolds number is not identical between the two studies, being $Re_\tau=200$ in the plane channel DNS and $Re_\tau \approx 175$ in the circular pipe experiment; the same applies to the forcing amplitude, $A^+=12$ in DNS and $A^+=13.8$ in the experiment; the periodic boundary conditions employed in the DNS are not fully equivalent to the actual flow conditions at the inflow and outflow sections of a finite length of a pipe, implying that a temporal transient in the former can easily be discarded, whereas the equivalent spatial transient in the experiment cannot be eliminated and affects the measurement.
However, the major difference --- and the one the present work sets out to investigate --- consists in the nature of the spatially distributed forcing applied at the wall: it is a sinusoidal function in the DNS study, where Eq.\eqref{eq:StTW} is enforced as a boundary condition, whereas in the experiment it consists of a piecewise-constant and time-periodic function.
The importance of waveform discretization was already discussed by ABBCQ. They mentioned how the wave realized with two segments only, i.e. $s=2$, is a limiting case of the discrete travelling wave (DTW), in which there can be no travelling direction for the wave, which becomes standing and works in a region of the parameter space where performance is suboptimal, especially in terms of energy saving. Moreover, the harmonic content of DTW was suggested to potentially explain some features of the experimental results shown in figure \ref{fig:DTW}. 

The importance of an in-depth understanding of discretization effects descends from the necessity of any experimental realization of such forcing to be, to an extent, spatially discrete. For temporal discretization (which is not particularly critical in experiments), these effects were systematically studied by \cite{cimarelli-etal-2013}, who considered various temporal waveforms to implement the spatially uniform spanwise-oscillating walls, and concluded that the sinusoidal waveform is the best overall in terms of energy savings.
Spatially discrete forcing, instead, has been rarely used so far and never discussed in terms of discretization effects.
\cite{kiesow-plesniak-2003} experimentally generated a localized crossflow in a three-dimensional turbulent boundary layer using a transverse running belt. 
In a numerical study, \cite{mishra-skote-2015} used only the positive cycle of steady square waves to enforce drag reduction in a turbulent boundary layer, under the rationale that similar performance compared to the complete control could be obtained with lower energy consumption.
\cite{straub-etal-2017} studied the spanwise oscillating wall technique in a low-aspect-ratio duct, and considered how it performs when only a fraction of the available surface is actuated. 
\cite{benard-etal-2021} dealt with the issue that several DBD actuators must be placed side by side to achieve a near-wall distribution of velocity that should be as spanwise uniform as possible.

The present paper describes the replica (including, first of all, the spatially discrete forcing) of the ABBCQ pipe flow experiment, based on direct numerical simulations (DNS), with a focus on drag reduction and energy efficiency. 
Not much information is available for spanwise forcing applied to cylindrical geometries. Besides the few early works \citep{orlandi-fatica-1997, quadrio-sibilla-2000, nikitin-2000} which proved with DNS the validity of spanwise (azimuthal) forcing in a pipe flow, only recently \cite{liu-etal-2022} tested spatially non-uniform spanwise forcing (in the form of stationary waves) in a pipe flow. Strictly speaking, thus, StTW for drag reduction have never been tested numerically in a pipe. For that purpose, in our DNS study we employ an original DNS solver, designed for efficient simulations of high-Reynolds number turbulent pipe flows, which addresses the problem of excess azimuthal resolution near the pipe axis: it exploits a  mixed discretization (Fourier in the homogeneous directions and compact finite differences in the radial direction) to decrease the azimuthal resolution as the pipe axis is approached.

The structure of the paper is as follows. In \S\ref{sec:methods} the DNS code will be presented and validated, the DTW forcing will be introduced, and the computational procedures adopted for the present study will be described. Results of the simulation campaign will be presented in \S\ref{sec:results}, touching upon drag reduction, power budget, and flow statistics. Discussion of the discretization effects and a critical evaluation of the present results against the ABBCQ experiment are contained in \S\ref{sec:discussion}; summarizing conclusions are provided in \S\ref{sec:conclusions}.

\section{Methods}
\label{sec:methods}

\subsection{The DNS code}

We solve by direct numerical simulation (DNS) the incompressible Navier--Stokes equations, written in non-dimensional form and cylindrical coordinates, for the primitive variables pressure $p$ and velocity $\vect{u}$. 
The axial, radial, and azimuthal directions are indicated with $x$, $r$, and $\theta$; the respective velocity components are $u$, $v$ and $w$. The axial length of the computational domain is $L$, the pipe radius is $R$; the complete azimuthal extent of $2 \pi$ is considered. 





Temporal discretization of the equations follows the usual partially-implicit approach in DNS of wall-bounded flows: the code implements a combination of the implicit Crank-Nicholson scheme for the viscous terms with a library of explicit schemes for the non-linear convective terms. In this work, the three-substeps, low-storage Runge--Kutta scheme described in \cite{rai-moin-1991} is used.

Spatial discretization, instead, deserves a specific discussion. The discretization is mixed, as in the Cartesian DNS solver \citep{luchini-quadrio-2006} which has inspired the present code. 
The homogeneous directions $x$ and $\theta$ call for a spectral discretization, naturally enforcing the required periodic boundary conditions with the computational efficiency of the pseudospectral approach. Compact, fourth-order accurate finite differences are used to discretize differential operators in the radial direction. 
The mixed discretization is at the root of an interesting feature of the present code, which addresses a fundamental problem with the DNS of the turbulent pipe flow.
Once the number of azimuthal Fourier modes is set according to physics-based considerations to yield the adequate azimuthal spatial resolution at the pipe wall, the azimuthal resolution increases above the required level as the pipe axis is approached. Such excess resolution not only is useless and thus a waste by itself, but also causes a steep rise in the computational cost of the simulation, since a vanishing spanwise cell size implies a vanishing time step, as the stability of the explicit temporal scheme requires that the Courant--Friedrichs--Lewy (CFL) number remains below the threshold dictated by the time integration scheme. Thus, while in general the computational cost of a DNS quickly grows with $Re$, for a pipe flow with a Fourier azimuthal discretization the cost has an even steeper rise because of the rapidly shrinking time step.

To handle this problem, short of accepting the vanishingly small time step, recent high-$Re$ simulations of pipe flow \citep[e.g.][]{pirozzoli-etal-2021} resort to an implicit treatment of the azimuthal convection terms. The alternative approach that is followed here derives from and extends the one introduced two decades ago by \cite{quadrio-luchini-2002}, who developed a DNS solver for the incompressible Navier--Stokes equations written in velocity-vorticity form, and used it for the DNS of the turbulent flow in an annular pipe.
Although the present code solves the Navier--Stokes equations in their primitive-variables formulation with a monolithic approach, in both cases the truncation of the azimuthal Fourier series changes with the radial coordinate, in such a way that the actual azimuthal resolution remains approximately constant across the pipe. This can be achieved smoothly and without interpolation, provided the spanwise Fourier series is combined, as in the present case, with a collocated method for the discretization of the radial direction. 

The radially varying number of azimuthal modes implies that some modes exist at the pipe wall at $r=R$ but do not reach the pipe axis at $r=0$: hence, these modes end in the bulk of the flow, where a suitable boundary condition has to be provided for them. On the ground that a well-designed DNS neglects, i.e. puts to zero, all the modes above its maximum wavenumber, in their original formulation \cite{quadrio-luchini-2002} used a simple homogeneous boundary condition also for these modes terminating in the fluid. 
Here, we take one step further and use as a boundary condition the same regularity conditions that are employed at the pipe axis. There, as shown for example by \cite{lewis-bellan-1990}, scalar and vector quantities require a different treatment. Let $\beta$ be the spanwise wavenumber, and let the superscript $\hat{\cdot}$ indicate the Fourier coefficient of a variable. For pressure, $\hat{p}(r) \sim r^{|\beta|}$ as $r \rightarrow 0$. For the velocity vector, the equivalent condition becomes:
\begin{equation}
   \begin{cases}
   \hat{u}(r) \sim r^{|\beta|}\\
   \hat{v}(r) \sim r^{|\beta|-1}\\
   \hat{w}(r) \sim r^{|\beta|-1}
   \end{cases}
    \text{when} \; \beta \neq 0 \;
    \qquad \qquad
      \begin{cases}
   \hat{u}(r) \sim r^{0}\\
   \hat{v}(r) \sim r\\
   \hat{w}(r) \sim r
   \end{cases}
   \text{when} \;  \beta = 0.
\label{eq:reg2}
\end{equation}

These regularity conditions are enforced not only for those standard modes that span the full radial extent of the pipe $0<r<R$ but also for those terminating in the bulk; this provides them with a smooth decay in the radial direction. Of course, as long as the baseline resolution is well chosen, no differences are expected in flow statistics between the present approach and the one employed by \cite{quadrio-luchini-2002}. 

The variable-modes approach can be easily programmed in a simple and general way. In the code, written in the CPL language \citep{cpl-website, luchini-2021}, a two-dimensional (for generality) array of pointers is used to reference into a variable-sized one-dimensional array, storing all the nonzero coefficients as a function of $r$. Although this exceeds the scope of this work, such a programming approach makes it straightforward to extend the use of variable modes: for example, in the plane channel flow it could be advantageous to have the number of both streamwise and spanwise modes become a function of the wall-normal coordinate.

As its Cartesian counterpart, the code is capable of parallel computing. It sports at the same time a shared-memory and a distributed-memory algorithm where a low-level message-passing strategy is employed on a computational domain subdivided into wall-parallel slices so that the two-dimensional inverse/direct FFTs can be computed locally to each machine. A standard message-passing version based on the MPI library is also available.

\subsection{The discrete travelling waves (DTW)}

\begin{figure}
\centering
\includegraphics[]{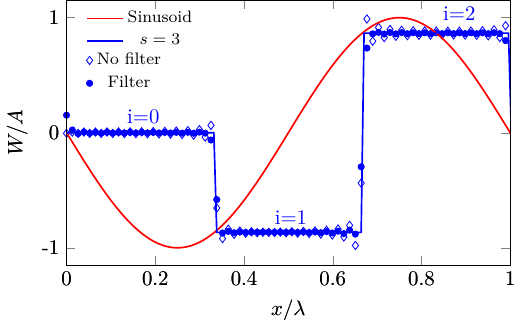}
\caption{Schematics of the spatial waveform for the control: the red line is the sinusoidal wave, and the blue line is its discrete approximation by three segments per period, i.e., $s=3$. The corresponding representation after the Fourier transform without (empty symbols) and with (filled symbols) spatial filtering is also depicted.}
\label{fig:gibbscontrol}
\end{figure}

The piecewise-continuous wall forcing shown with a blue line in figure \ref{fig:DTW} does not lend itself to an immediate description within a Fourier discretization. An equivalent problem was faced by e.g. \cite{ricco-hahn-2013} and \cite{wise-ricco-2014}, who described spatially discontinuous forcing distributions with a Fourier discretization. 
\cite{mishra-skote-2015}, who dealt with the same issue, reported that oscillations introduced at the discontinuities (the so-called Gibbs phenomenon) produce instabilities and large numerical errors, and therefore need to be adequately treated. 

The present DTW forcing, shown in figure \ref{fig:gibbscontrol}, is defined by a number $s$ of segments discretizing the continuous sinusoidal counterpart. 
One wavelength of the piecewise-constant travelling wave is written analytically as:
\begin{equation}
w(x,t;s)= A \sin \Bigl( \omega t - \frac{2\pi i}{s} \Bigr) \qquad 
   \frac{i}{s} \lambda \leq x < \frac{i+1}{s} \lambda
\label{eq:DTW}
\end{equation}
where $\lambda = 2 \pi / \kappa_x$ stands for the wavelength of the wave, and $i$ is an integer spanning the range $0 \leq i < s-1$ (note that this definition is non-unique, as the phase difference between the DTW and the sinusoid could be chosen differently without altering the results discussed in the following). The waveform is then periodically extended to the whole axial length of the pipe, which always fits an integer number of wavelengths.

To avoid the appearance of spurious oscillations, the discontinuities in the DTW are regularized via a smoothing Gaussian filter. The expression of the filter in physical space is:
\begin{equation}
G(x) = \left( \frac{6}{\pi\Delta^2} \right)^{1/2} \exp \Bigl( -\frac{6 x^2}{\Delta^2} \Bigr)
\label{eq:filter}
\end{equation}
with $\Delta$ the filter width; the filtered wall forcing is obtained via convolution of the original one with the kernel $G(x)$ in physical space, and then Fourier-transformed.
The filter width was carefully determined after a parametric study in the preliminary work carried out by \cite{biggi-2012}; the value of $\Delta$ is chosen as to strike a compromise between the range of scales/positions affected by the filter and the magnitude of the residual oscillations.  
The employed filtering is represented in figure \ref{fig:gibbscontrol}, where it can be appreciated that oscillations are reduced to a very small level, while the step function remains relatively sharp.

\subsection{Validation}
\label{sec:validation}

\begin{figure}
\centering
\includegraphics[]{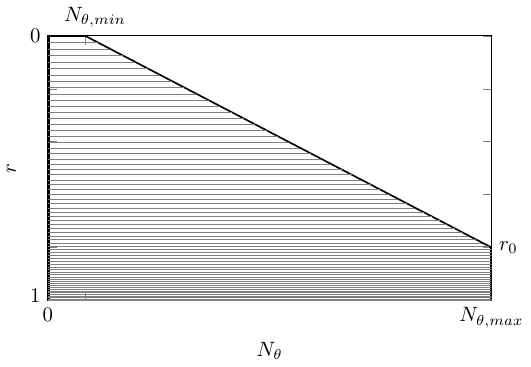}
\caption{Schematic representation of the radially-varying number of azimuthal nodes (only the positive ones are shown). The maximum value $N_{\theta,max}$ is constant from the pipe wall at $r=1$ to the radial position $r=r_0=0.8R$, then decreases linearly to the value $N_{\theta,min}$ at the pipe axis. In this plot, as in the main simulations, $N_r=100$ and every other radial point is omitted for clarity.}
\label{fig:variablemodes}
\end{figure}

Before delving into the actual study, we preliminarily assess, for a canonical turbulent pipe flow, to what extent the radially varying number of azimuthal modes affects the solution and the computational efficiency of the code. 
For a turbulent pipe flow at a bulk Reynolds number of $Re_b = U_b D / \nu = 4900$, which is the value used for the rest of the study, two configurations are considered: one in which the spanwise truncation of azimuthal modes is simply kept constant at the value $N_{\theta,max}$, and the other in which the azimuthal modes decrease with $r$ from $N_{\theta,max}$ to $N_{\theta,min}$. 
As shown schematically in figure \ref{fig:variablemodes}, based on previous experience, we have set up the azimuthal discretization such that the maximum resolution with $N_{\theta,max}=96$ (i.e. 193 azimuthal modes, from $-N_{\theta,max}$ to $+N_{\theta,max}$) holds in a near-wall layer $r_0 \leq r \leq R$, with $r_0=0.8R$; for larger wall distances, the number of azimuthal modes decreases linearly with $r$ so that $N_{\theta,min}=4$ (i.e. 9 modes, from $-N_{\theta,min}$ to $+N_{\theta,min}$) at the centerline. The discretization also employs 384 streamwise modes and 100 radial points, for a pipe with a length of $L=22R$. 

\begin{figure}
\centering
\includegraphics[width=0.75\textwidth,]{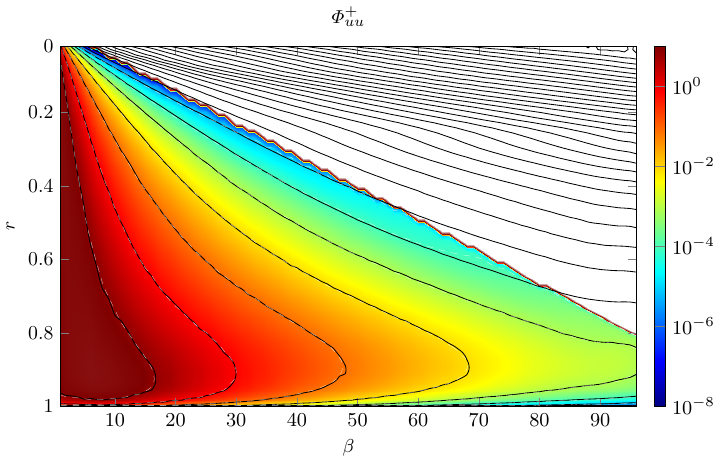}   
\caption{Spectral energy density $\Phi^+_{uu}$ of the axial velocity fluctuations in the $r-\beta$ plane. Black contour lines are from the simulation with constant azimuthal resolution, and the colormap with white dashed contours is from the the simulation where the number of azimuthal modes varies with $r$. Contour levels start from $10^0$, spaced by one order of magnitude.}
\label{fig:sp_zy}
\end{figure}

In figure \ref{fig:sp_zy} we compare the spectral energy density $\Phi_{uu}^+(\beta)$ of the streamwise velocity fluctuations in the $r-\beta$ plane, as computed from the two simulations. As expected, the simulation with the standard discretization has an energy density (black contour lines) that peaks near the wall; toward the axis of the pipe, it decreases to very low values already at a rather small azimuthal wavenumber. For the considered spanwise resolution, the near-wall maximum captured at the largest $\beta$ is about $10^{-4}$; as the pipe axis is approached, the energy levels decrease, and at the axis, the energy levels drop below $10^{-30}$ for the largest $\beta$. Once a radially varying number of azimuthal modes is employed (colored contour), this waste is avoided, and densities below $10^{-7}$ are not computed in the core region of the pipe. Importantly, the white dashed contour lines of the variable-modes simulation demonstrate that the two datasets overlap perfectly in the region of the $r-\beta$ plane where they are both defined. 

The two cases produce a virtually identical value of the friction coefficient, namely $C_f=9.52 \cdot 10^{-3}$, which is very near to the value of $9.45 \cdot 10^{-3}$ predicted by the Blasius power-law $C_f=0.0791Re_b^{-1/4}$ \citep{schlichting-gersten-2000}: a more than satisfactory agreement, given the known difficulties for such correlations at low values of the Reynolds number.  

An additional validation step is carried out in the presence of the spanwise forcing. The availability of the recent work by \cite{liu-etal-2022}, who numerically tested the drag reduction capabilities of standing waves, allows us to repeat a representative set of their simulations for comparison. As in that work, a fixed $Re_\tau=180$ is enforced, and the pipe length is $L=6 \pi R$. We keep the same spatial resolution employed above since it is nearly identical to the one employed in their study. We replicate one of their cases at $A^+=12$ and $\lambda^+=424$ with 40.4\% drag reduction, obtaining 40.6\% drag reduction. Another case with $A^+=6$ and $\lambda^+=1695$ is reported to yield 28.1\% drag reduction; here it yields 28.5\%. Finally, the case with $A^+=30$ and $\lambda^+=1695$ is confirmed to lead to a complete relaminarization of the flow.

In terms of computational efficiency, the code has been tested on an Intel Cascade Lake 8260 processor. The single-core solution of one Runge--Kutta time step (i.e. the sum of the three substeps) requires $\simeq 20$ seconds when the full azimuthal nodes are retained, and $\simeq 15$ seconds with the variable modes set up as described above, with a 25\% savings in computing time. However, as previously discussed, the true advantage of employing a radially varying number of azimuthal modes lies in the larger time step size allowed by the stability condition. In these tests, the same value of (unitary) CFL corresponds in the first case to a time step of $\Delta t^+\simeq 0.008$, which becomes in the second $\Delta t^+\simeq 0.08$, demonstrating one order of magnitude larger time steps when the azimuthal modes are truncated.
Finally, the usage of variable modes is beneficial also for the memory occupation of the code, which in this configuration (i.e. with a Runge--Kutta method that stores the solution at one previous time level) amounts to 331 MB of RAM instead of 434 MB when $N_\theta$ is kept constant at $N_{\theta,max}$ across the pipe.

\subsection{Computational procedures}

The numerical study described in the remainder of the paper is carried out at the same Reynolds number of the ABBCQ experiment, namely $Re_b \equiv U_b D / \nu = 4900$, where $U_b$ is the bulk velocity, and $D=2R$ is the pipe diameter. Every simulation is performed by adjusting at every time step the axial homogeneous pressure gradient in such a way that the flow rate does not vary in time, i.e. by following the Constant Flow Rate (CFR) strategy, as defined by \cite{quadrio-frohnapfel-hasegawa-2016}. 
An uncontrolled simulation serves as the reference case and establishes the corresponding nominal values for the drag and the friction-based Reynolds number, $Re_\tau \approx 170$.

The length of the computational domain is $22R$; $384 \times 192$ Fourier modes are used to discretize the streamwise and azimuthal directions respectively, while $N_r = 100$ nodes discretize the radial direction.
Once the additional modes used to remove the aliasing error are accounted for, the spatial resolution of the reference case is $\Delta x^+=4.8$ and $r \Delta  \theta^+=2.8$ at the pipe wall. The radial resolution varies from $\Delta r^+=0.7$ near the wall to $\Delta r^+=2.4$ at the centerline. The radially-varying modes are set up as described above in \S\ \ref{sec:validation}, with a linear variation from $N_{\theta,min}=4$ to $N_{\theta,max}=96$, and $r_0=0.8R$. The time step is dynamically adjusted to satisfy the constraint CFL=1.

The reference experimental dataset by ABBCQ, already shown in figure \ref{fig:DTW}, consists of data taken at various oscillation frequencies, ranging from $\omega^+=-0.25$ to $\omega^+=0.2$ (note that quantities indicated with the + superscript are defined in terms of the friction velocity of the uncontrolled flow). The forcing amplitude is also fixed at $A/U_b=1$ (or $A^+ \approx 14$). 
Owing to the fixed size of the device producing the DTW, the wavelength could not be changed continuously in the experiment. For the DTW produced with three moving segments, $s=3$, it was $\lambda / R = 4.38$ (equivalently, the streamwise wavenumber was $\kappa_x R=1.43$ or $\kappa_x^+=0.0082$).

The present numerical experiments include three sets of simulations.
In one, labeled SIN, the idealized sinusoidal boundary condition \eqref{eq:StTW} is applied, while the second and third sets consider DTW, realized with a relatively fine ($s=6$, case S6) and coarse ($s=3$, case S3) discretization of the waveform, according to Eq.\eqref{eq:DTW}. In particular, case S3 has the closest correspondence to the experiment. The wavelength and amplitude of the forcing are nominally identical to those of the experiments, whereas the oscillation frequency is varied with a relatively fine step, and covers a slightly larger range.

Twenty-nine simulations are performed for each forcing type, for a total of 88 cases including the reference one. Understanding the harmonic content of the forcing led us to run 45 additional simulations where only some harmonics of the DTW are included. 
Each case is run for a total of 1000 convective time units; the first half of the time history is discarded to avoid the influence of the initial transient, which can be particularly long at times, especially for the cases with the highest drag reduction. 
In a few selected cases, a database is saved for later statistical analysis; for these, 50 oscillation cycles are sampled, and four phases are stored for each cycle. 
Most of the simulations are run serially (i.e. on one core each) on a single machine equipped with an Intel Xeon Phi processor and 68 cores, which was kept busy for approximately two months.	

\section{Results}
\label{sec:results}

\subsection{Drag reduction}
\label{sec:DR}

\begin{figure}
\centering
\includegraphics[scale=1]{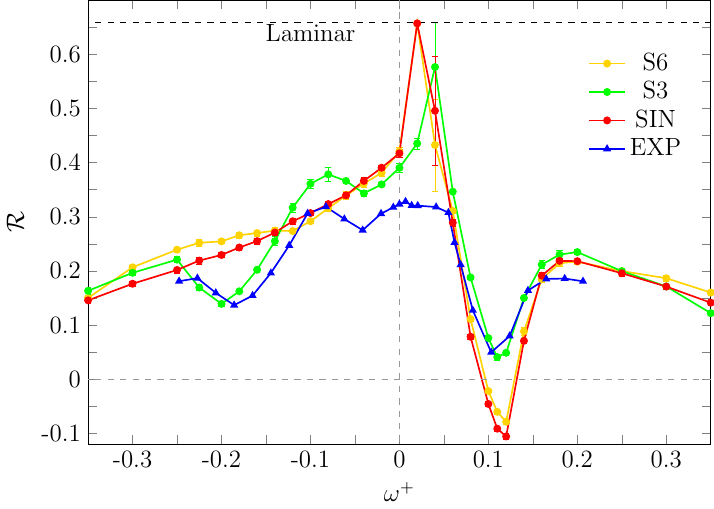}
\caption{Drag reduction rate $\DR$ computed with DNS against experimental measurements from ABBCQ (blue triangles, blue line). The plot shows the laminar limit (upper dashed line), and quantifies with bars the error deriving from the finite averaging time (see text).}
\label{fig:RDisc}
\end{figure}

We start by looking first at the raw changes in skin-friction drag. The drag reduction rate is:
\begin{equation}
\DR= 1 - \frac{C_f}{C_{f,0}} = 1 - \frac{\tau_w}{\tau_{w,0}}
\label{eq:R}
\end{equation}
where the subscript 0 indicates quantities evaluated for the uncontrolled flow, $C_f$ is the friction coefficient, defined in the usual way as:
\[
C_f = \frac{2 \tau_{w,x}}{\rho U_b^2} ,
\]
and the last equality in Eq.\eqref{eq:R} only holds for CFR simulations. Note that, in cylindrical coordinates, the longitudinal and azimuthal components of the mean wall shear stress are:
\[
\tau_{w,x} = - \mu \frac{\partial \overline{u} }{\partial r} \Bigr|_{r=R}, \qquad 
\tau_{w,\theta} = - \mu \left( \frac{\partial \overline{w} }{\partial r} - \frac{\overline{w} }{R} \right) \Bigr|_{r=R} .
\]
where the overbar indicates temporal average.

Figure \ref{fig:RDisc} compares the output of our simulations with the ABBCQ measurements for $s=3$. The first striking observation is that the experimental data, labelled as EXP, are quite far from the results obtained with the ideal sinusoidal forcing, case SIN. 
The qualitative look of the curves is similar. The sudden shift from drag reduction to drag increase for waves travelling at a phase speed comparable to the convection speed of the near-wall turbulent structures known to take place in the planar caase \citep{quadrio-ricco-viotti-2009} is confirmed. However, significant quantitative differences do exist. 
The maximum drag reduction obtained with SIN is about twice the experimental one, and peaks at 66\%, which at this $Re$ corresponds to full relaminarization of the flow. Moreover, the SIN data do not present the evident wiggles of the experimental $\DR=\DR(\omega^+)$ curve. 
The error bars plotted in figure \ref{fig:RDisc} refer to the finite averaging time, and are computed according to the procedure introduced by \cite{russo-luchini-2017}. They are generally small, and confirm the deterministic nature of the wiggles, which are not an artifact of the measurement procedure. An evident exception are the points at $\omega^+=0.04$, where the flow is on the verge of relaminarisation, and alternately visits a turbulent state and a nearly laminar state, switching between them over a long time scale. This observation also explains the apparent disagreement of the curves SIN, S3 and S6 at this very specific control point.
 
Such differences between the experimental data and the expected drag reduction from plane channel DNS were already noticed in the original paper by \cite{auteri-etal-2010}, in a comparison with plane-channel information at $Re_\tau=200$, but their identification and interpretation are now easier. The one-to-one comparison made possible by the present data rules out several alternate explanations for these discrepancies, as for example the difference between circular and planar geometries (which, in fact, acts to actually reduce differences), or the slightly different parameters of the forcing.

Case S3 (green curve) is the one that should most closely correspond to EXP (blue curve); indeed, experiments and simulations, while not overlapping, do show much better agreement. Both datasets present, especially at negative frequencies, the large wiggles that are missing in the SIN curve. Notably, at certain frequencies, the discrete wave realized by S3 achieves a distinctly larger $\DR$ than SIN, a feature that was not previously observed. Relaminarization is only partial for S3, with a maximum $\DR$ of 57\%, obtained at a frequency slightly larger than the optimum for SIN; as for EXP, the region of drag increase at $\omega^+ \approx 0.1$ is correctly identified, but $\DR$ remains positive and no drag increase is measured.

Case S6 implements the same travelling wave tested in experiments, but yields a better approximation of the sinusoidal waveform. Drag reduction data from S6 resemble very much those from SIN, and full laminarization is achieved at the smallest positive frequency, indirectly confirming the key role of control discretization. Still, a slightly diminished drag increase and the presence of wiggles (albeit of smaller amplitude) indicate that discretization effects remain at work even in the S6 case.
Reasons explaining the observed discrepancies among these data sets will be discussed later in \S\ref{sec:discussion}.

\subsection{Power budget}

\begin{figure}
\centering
\includegraphics[width=\textwidth]{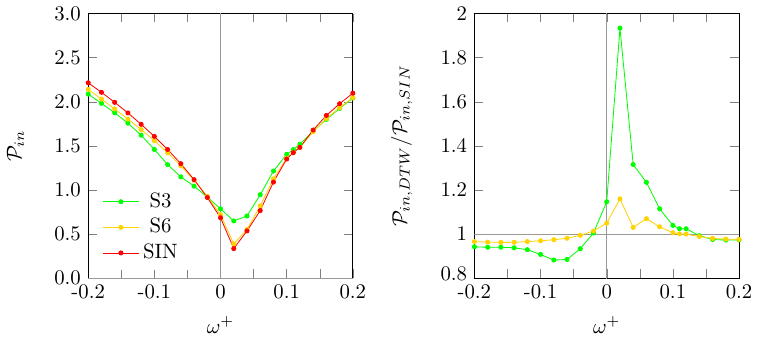}
\includegraphics[width=\textwidth]{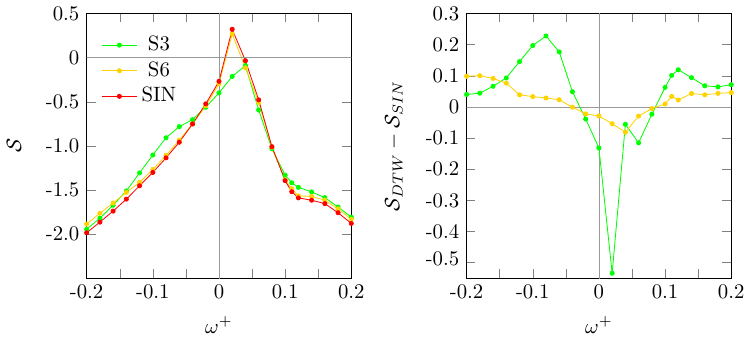}
\caption{Input power ratio $\Pin$ (top) and net savings $\NS$ (bottom). The right panels emphasize variations from the SIN case.}
\label{fig:PinAndS}
\end{figure}

With active control, it is important to complement the information regarding drag reduction with the cost of the input power, conveyed by the power ratio $\Pin$ between the power required to create the control action and the power $P_0$ per unit wall area required to drive the uncontrolled pipe flow. Hence, if saving energy is the ultimate interest, more than the drag reduction rate $\DR$ itself, the most informative quantity is the net power saving $\NS$, defined as:
\begin{equation}
\NS = \DR - \Pin .
\end{equation}

In the present application, the control acts in the azimuthal direction only, and the input power that an ideal control system transfers to the viscous fluid, normalized with the pumping power $P_0$, is:
\begin{equation}
\Pin = \frac{1}{P_0} \frac{1}{2 \pi R L T} \int_0^T \int_0^{2\pi} \int_0^L  
- \mu w \left( \frac{\partial w}{\partial r} - \frac{w}{R} \right) \Bigr|_{r=R} R \ dx \ d \theta \ dt .
\label{eq:Pin}
\end{equation}

Note that the term $w/R$ in the integrand of Eq.\eqref{eq:Pin} is sometimes incorrectly omitted in existing studies, but its present in the correct expression of the wall shear stress in cylindrical coordinates.  
Figure \ref{fig:PinAndS} (top) plots how $\Pin$ changes with the control frequency, in comparative form between SIN and cases S3, and S6. The outcome is not obvious.
The input power varies significantly with the control parameters; as in the planar case, it is minimum in the region (small positive frequencies) where the drag reduction is maximum. Of primary interest here, however, are the relative differences between the sinusoidal forcing and DTW, highlighted in the right panel. 
It is seen that at not-too-small negative frequencies (say $\omega^+<-0.02$) and large positive ones ($\omega^+>0.14$), DTW are less expensive than their sinusoidal counterpart, with larger differences, of the order of 10\%, observed for S3 and backward-travelling waves. However, the opposite scenario is observed at small positive frequencies, i.e. in the most interesting region where the control is supposed to work, owing to the lower absolute energetic costs. For example, at $\omega^+=0.02$ the power $\Pin$ required by S3 is twice that for SIN.


The combined dependence of both $\DR$ and $\Pin$ on the control parameters determines the changes to the net savings $\NS$, plotted in figure \ref{fig:PinAndS} (bottom).
The large differences in terms of relaminarization (or lack thereof) observed at the smallest frequencies for the various types of forcing blurs the picture further; moreover, the ratio of $\NS$ obtained with DTW and SIN becomes a delicate indicator whenever $\NS$ approaches zero. 
The most expensive S3 forcing shows significant extra savings compared to SIN; however, this happens for frequencies where $\NS$ is large and negative, which makes the improved performance pointless. Wherever $\NS$ is positive (or, as in this case at a relatively large forcing amplitude, wherever it approaches zero), S3 presents a significant efficiency gap compared to SIN. 
At any rate, in correspondence of the optimum parameters, SIN remains the best forcing type, in terms of both $\Pin$ and $\NS$.


\subsection{Flow statistics}

\begin{figure}
\centering
\includegraphics[]{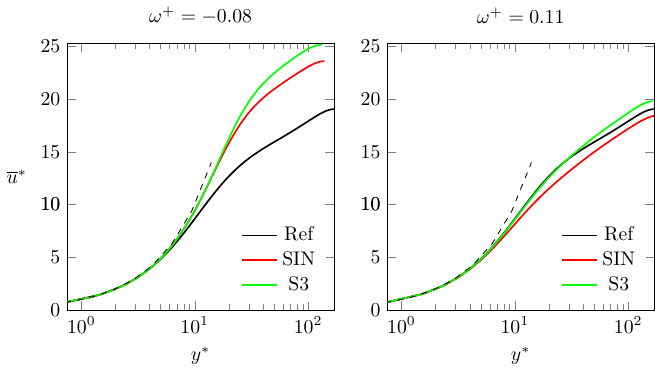}
\caption{Mean streamwise velocity profile for two cases with drag reduction ($\omega^+=-0.08$, left) and drag increase or minimal drag reduction ($\omega^+=0.11$, right). Cases SIN and S3 are compared to the uncontrolled flow.}
\label{fig:uMean}
\end{figure}

Turbulence statistics are inspected here for the sole purpose of verifying whether or not the discrete form of the forcing alters the flow significantly, besides the already quantified different level of $\DR$. To this aim, we focus on SIN and S3 and pick two cases: one is at $\omega^+=-0.08$ and consistently yields a large positive $\DR$, whereas the second is at $\omega^+=0.11$ and yields drag increase --- more precisely, a negative $\DR$ for SIN and a very small positive value of $\DR$ for S3. 
In figure \ref{fig:uMean}, the mean streamwise velocity profile $\overline{u}$ is plotted against the wall distance $y=1-r$ in the law-of-the-wall form by using the actual friction velocity as a reference velocity. This is the so-called true viscous scaling \citep{quadrio-2011}, indicated with an asterisk superscript. 
It can be appreciated that the profiles of $\overline{u}^*$ differ essentially only because of the different value of $\DR$, which translates into a different vertical shift of the logarithmic portion of the profile. 
Indeed, owing to the scaling employed, all profiles collapse in the viscous sublayer; the vertical shift $\Delta B^*$ appears in the logarithmic region (which is not particularly wide, since $Re$ is low and further lowered by drag reduction). As a proxy for the shift, we consider the value of $\overline{u}^*$ at $y^*=100$, and obtain a $\Delta u^*$ of 5.1 for SIN and 6.9 for S3 when $\omega^+=-0.08$, in agreement with the larger drag reduction of the latter case. For the positive frequency, instead, $\Delta u^*$ is $-0.8$ for SIN and $+0.9$ for S3, once again in agreement with the small drag increase of the former and the small drag reduction of the latter. Besides the different levels of $\DR$ and the consequent different vertical shifts, no other difference can be appreciated in the profiles.

\begin{figure}
\centering
\includegraphics[]{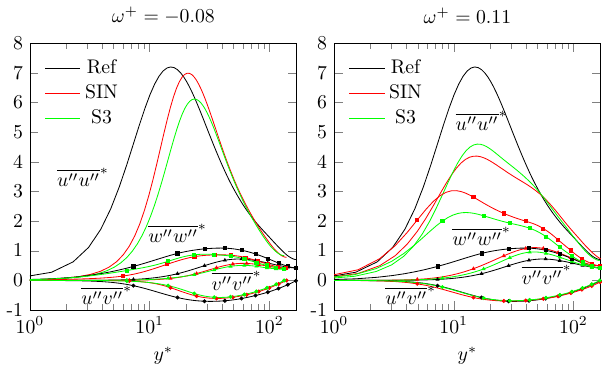}
\caption{Wall-normal profiles of the stochastic turbulent stresses. Colors as in figure \ref{fig:uMean}.}
\label{fig:rms}
\end{figure}

A similar picture emerges by looking at second-order statistics, i.e. for example the variance of velocity fluctuations. In the present case, the fluctuations should be defined by accounting not only for the mean flow but also for the control-induced coherent velocity field with zero average. To this purpose, we use a classic triple decomposition, where a generic quantity $a$ is decomposed as $a = \overline{a} + \tilde{a} + a''$, i.e. into its mean, coherent, and stochastic components. The sum of the coherent and stochastic components is indicated as $a'=\tilde{a}+a''$.
The mean component is obtained by averaging each quantity in time and along the azimuthal and streamwise directions; the coherent part, instead, derives from averaging together points at the same phase $\phi = \kappa_x x - \omega t$ after removal of the mean value. The difference between the instantaneous field and the corresponding mean and coherent parts defines the stochastic field.

In the present work, we do not consider the coherent components, since their magnitude is negligible compared to the stochastic ones. An exception is obviously the Stokes layer contributions $\tilde{w}$ and $\widetilde{ww}$, and a small streamwise modulation of the $\tilde{u}$ component, which becomes apparent in its wall derivative.
Figure \ref{fig:rms} depicts the wall-normal profiles of the (stochastic) Reynolds stresses normalized with the actual friction velocity, i.e. $\overline{u''u''}^*$, $\overline{v''v''}^*$, $\overline{w''w''}^*$, and $\overline{u''v''}^*$.   
The streamwise fluctuations always decrease with control, either continuous or discrete, for both drag reduction and drag increase. This can be attributed to the strengthened redistribution action of the pressure-strain term, which moves energy towards the spanwise and wall-normal fluctuations. 
The redistribution is enhanced by the tilting of the structures \citep{yakeno-hasegawa-kasagi-2014, gallorini-quadrio-gatti-2022}. For example, the maximum tilt angle, defined as in \cite{yakeno-hasegawa-kasagi-2014}, here is $14^\circ$ for $\omega^+=-0.08$, and becomes $34^\circ$ for $\omega^+=0.11$ with SIN and $\approx 27^\circ$ at the same frequency with S3.

For the negative drag-reducing frequency $\omega^+=-0.08$, the reduction in intensity is accompanied by a slight shift of the near-wall peak further from the wall. The shift is larger, and the decrease of the peak value is much larger, for S3 than for SIN. 
Since the different $\DR$ is already accounted for by the * scaling, the extra reduction of the peak value provided by S3 is attributed to the highly subcritical turbulent state (described later in \S\ref{sec:localized}) reached by the turbulent flow in this case.
The positive drag-increasing frequency $\omega^+=0.11$ has the location of the wall-normal peak approximately unchanged, but the profiles after the peak show a linear (in semi-logarithmic scale) region that is absent in the reference profile (at these $Re$). Such linear regions also appear in the profile for $\overline{w''w''}^*$, and resemble those observed by \cite{lee-moser-2015} in canonical channel flows, but at much higher Reynolds numbers ($Re_\tau > 550$).
It is worth mentioning that the shift of $\overline{u''u''}^*$ in the wall-normal direction qualitatively agrees with the observation of \cite{gallorini-quadrio-gatti-2022}, in which the authors pointed out a movement away from the wall of the quasi-streamwise vortices related to $\DR$ in case of drag reduction and minor modifications in case of drag increase.
The spanwise and wall-normal diagonal components of the Reynolds stress tensor show minor changes for the negative frequency, but very large increases for the positive frequency, suggesting lesser changes in the drag-reduced turbulent flow, and minor effects of the discrete forcing as long as it works to reduce drag, whereas drag increase implies important modifications also far from the wall.
The off-diagonal component $\overline{u''v''}^*$ presents changes that are expected from the FIK identity \cite{fukagata-iwamoto-kasagi-2002} (although contributions from additional terms are present due to streamwise inhomogeneity) and decreases whenever drag reduction is present.

\begin{figure}
\centering
\includegraphics[width=\textwidth]{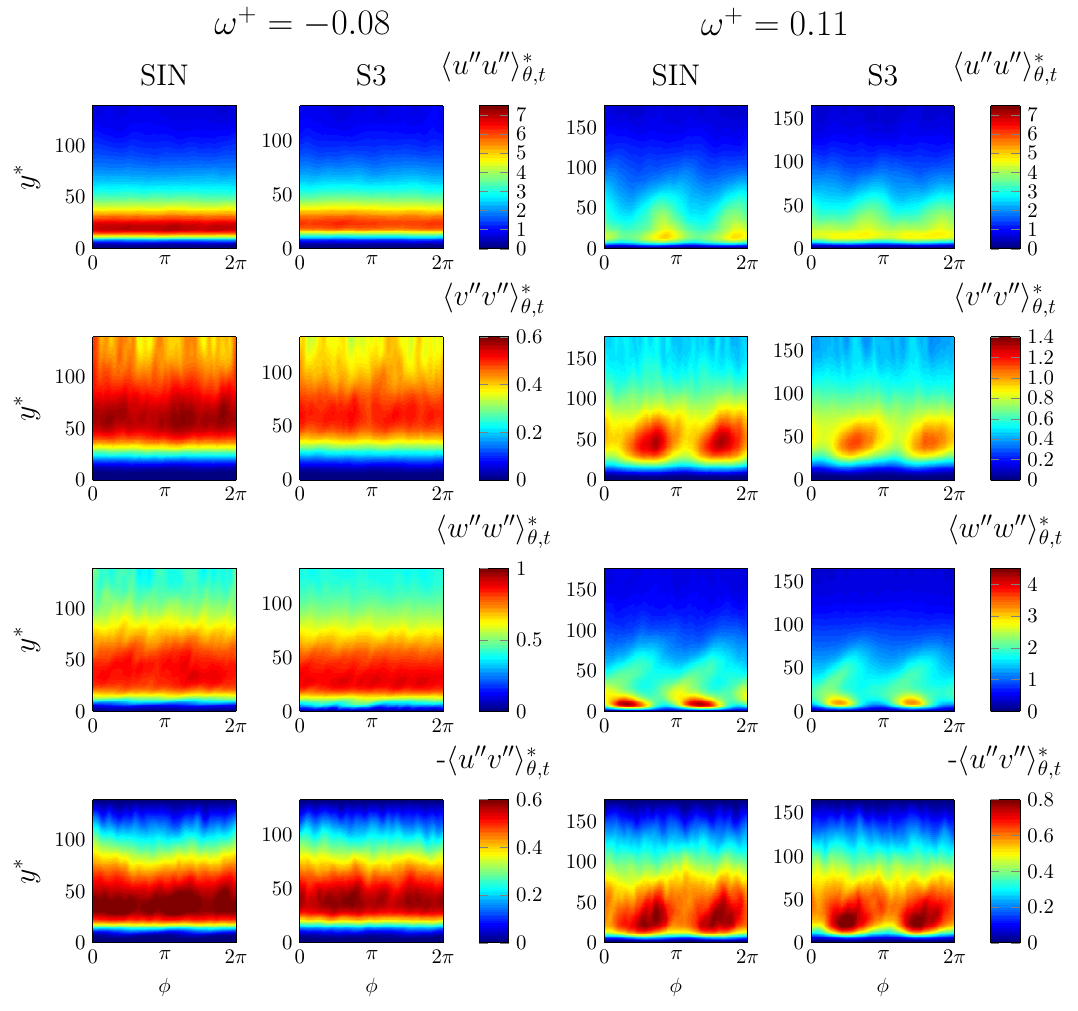}
\caption{Components of the stochastic Reynolds stress tensor as a function of the radial coordinate $r$ and the phase $\phi$, for two cases with drag reduction ($\omega^+=-0.08$, left) and drag increase or minimal drag reduction ($\omega^+ = 0.11$, right).}
\label{fig:RS}
\end{figure}

With a streamwise-varying forcing, the streamwise direction is not homogeneous anymore, and flow statistics may vary also with the phase $\phi = \kappa_x x - \omega t $ of the forcing. 
Figure \ref{fig:RS} plots the stochastic Reynolds stresses as a function of $y$ and $\phi$, after averaging in time and over the azimuthal direction. 
For the cases with drag reduction, as originally observed by \cite{quadrio-ricco-viotti-2009}, no streamwise modulation of the statistics is observed, if not for some residual statistical noise: the flow does not get directly altered along the forcing wavelength, and its main change is the reduced level of wall friction. 
The cases with drag increase, instead, do show a strong modulation over the forcing wavelength, which extends quite far from the wall. As in the planar case, the periodic modulation possesses a structure in the radial direction that is influenced by the specific spatial shape of the coherent generalized Stokes layer \citep{quadrio-ricco-2011}.  
The modulation is interpreted as a sort of resonance between the convection speed of the near-wall turbulence structures and the phase speed $\omega / \kappa_x$ of the travelling wave.
It is worth noting that such modulation is visible for S3 too, although $\DR$ is slightly positive. The coexistence of (small) drag reduction and a streamwise-modulated flow can be attributed to the complex interactions taking place with DTW: as shown later in \S\ref{sec:discretization}, the superposition of different harmonics is such that, in this case, features of drag-reducing and drag-increasing flows are observed simultaneously.

\section{Discussion}
\label{sec:discussion}

Replicating the ABBCQ experiment by DNS has confirmed that some features of the experimental data derive from the different nature of continuous and discrete forcing. 
In this Section, we discuss the reason for the two major differences: the wiggles in the curve $\DR = \DR(\omega^+)$ (which are always present for discrete forcing, regardless of the nature of the measurement), and the larger maximum drag reduction observed in the simulations.

\subsection{The role of different harmonics}
\label{sec:discretization}

Although in both the continuous and discrete case the spanwise wall velocity $W$ varies in time between $\pm A$, a meaningful comparison between DTW and SIN at the same amplitude is not obvious. 

\begin{figure}
\centering
\includegraphics[width=\textwidth]{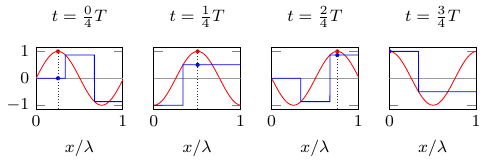}
\caption{SIN (red line) and S3 (blue line) wall forcing along one wavelength, at four different times during the period $T$. At each time, the dots mark the position at which SIN is maximum.}
\label{fig:controlt4}
\end{figure}

It is easy to show that the staircase function sketched in figure \ref{fig:DTW} and expressed analytically by Eq.\eqref{eq:DTW} has a lower amplitude compared to the sinusoidal forcing once averaged over the forcing period. 
Figure \ref{fig:controlt4} plots SIN and S3 at four different instants during one period. As time progresses, the sinusoidal wave is simply shifted in space, whereas DTW sees its waveform modified during the period: by focusing on a specific phase $\phi = \kappa_x x - \omega t $ (for example, as in the figure, where SIN has its maximum), it can be seen that DTW assumes different values at different times. 
For the generic phase $\phi$, the average intensity of DTW can be quantified analytically. By defining $\Delta_s = 2 \pi / s$ as the width of the each constant piece of the DTW, $W(\phi)$ can assume any value from $A\sin(\phi-\Delta_s/2)$ to $A\sin(\phi+\Delta_s/2)$. 
Taking the average of the sinusoidal function over the interval $\Delta_s$, one obtains:

\begin{equation}
W(\phi) = \frac{1}{\Delta_s} \int_{\phi-\Delta_s/2}^{\phi+\Delta_s/2} A \sin(\phi') \, d\phi' = -\frac{A}{\Delta_s}\Bigl[ \cos(\phi+\frac{\Delta_s}{2}) - \cos(\phi-\frac{\Delta_s}{2}) \Bigr].
\end{equation}

The previous equation can be rearranged, with the aid of the prosthaphaeresis formulae, into the following expression:
\begin{equation}
W (\phi) = 2 \frac{A}{\Delta_s} \sin \left( \frac{\Delta_s}{2} \right) \sin(\phi) = 
\underbrace{\frac{s}{\pi} \sin \Bigl( \frac{\pi}{s} \Bigr)}_{\gamma(s)} A \sin(\phi) .
\label{eq:paDTW}
\end{equation}

Hence, the averaged DTW can be written as $\gamma(s) A \sin (\phi)$, i.e. proportional to $A \sin(\phi)$, but its effective amplitude is equal to $A$ only for the ideal sinusoidal forcing described by Eq.\eqref{eq:StTW}, for which $s \to \infty$ and $ \lim_{s \to \infty} \gamma(s) = 1 $. 
Compared to the nominal amplitude $A$, the amplitude of DTW is decreased by a factor $\gamma(s)$, which depends on the number of slabs discretizing the sinusoid. For the S3 case, $\gamma(3) = 0.83$. This is consistent with the S3 forcing not reaching relaminarization, as shown in figure \ref{fig:RDisc}, because of its smaller effective amplitude. In fact, we have verified with an additional simulation, run with sinusoidal forcing for $\omega^+=0.02$ and a forcing intensity reduced by a factor 0.83, that the flow does not reach the fully laminar state, and yields $\DR=0.43$.

However, this is only part of the whole picture. Once the importance of discretization is recognized, it is necessary to proceed further to properly describe DTW: in fact, the reduced amplitude discussed above cannot explain, for example, other results reported in figure \ref{fig:RDisc}, where in the range $-0.12 \leq \omega^+ \leq -0.06$ drag reduction due to the discrete S3 forcing is {\em larger} than that achieved by SIN, while the reduced effective amplitude discussed above would suggest otherwise. 

The DTW described by Eq.\eqref{eq:DTW} can be expanded into an infinite Fourier series. Following ABBCQ, a DTW characterized by its three parameters $A$, $\kappa_x$ and $\omega$, and discretized into $s$ segments, is rewritten as the following sum:
\begin{equation}
\begin{aligned}
W(x,t;s) = & A \sum_{m=0}^{\infty}
\frac{\sin[ (ms +1) \pi/s ]}{(ms +1)\pi/s} \sin[\omega t - \kappa_x(ms +1)x]+ 
\\
& + \frac{\sin[ ((m+1)s -1)\pi/s ]}{((m+1)s -1)\pi/s}\sin[\omega t + \kappa_x((m+1)s -1)x] .
\end{aligned}
\label{eq:DTWfourier}
\end{equation}

\begin{figure}
\centering
\includegraphics[width=\textwidth]{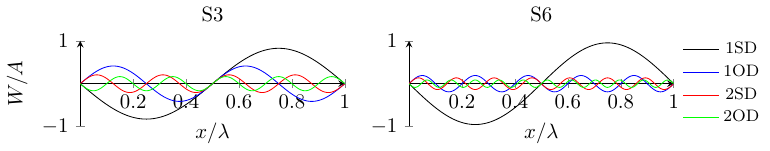}
\caption{Sketch of the first two pairs ($m=0$ and $m=1$) of harmonics for S3 (left) and S6 (right). Each pair is made by two waves: one travelling in the same direction (1SD, 2SD) and the other in the opposite direction (1OD, 2OD) of the nominal DTW.}
\label{fig:sines}
\end{figure}

The DTW written as in \eqref{eq:DTWfourier} is made by two families of sinusoidal waves: one family has a phase speed of the same sign of the nominal wave (and thus travels in the same direction, SD), while the other has the opposite sign and travels in the opposite direction (OD). 
The temporal frequency of the waves is unchanged (in absolute value) and equals the nominal one, but the effective amplitude (always $\le A$) and streamwise wavenumber (always $\ge \kappa_x$) of each wave depend on the number of slabs $s$ and on the index $m$ of the series. The amplitude of the harmonics decreases with $m$, and the wavenumber correspondingly increases.
The averaged DTW derived above in Eq.\eqref{eq:paDTW} is recognized as the first SD wave with $m=0$. 
Figure \ref{fig:sines} plots as an example the first ($m=0$) pair (1SD and 1OD) and the second ($m=1$) pair (2SD and 2OD), for a visual appreciation of the relative amplitude and wavelength. 

\begin{table}
\centering
\begin{tabular}{l|c|ccccc|ccccc}
Case             & SIN      & S3       & 1SD       & 1OD       & 1p       & 1p+2p    & S6       & 1SD      & 1OD      & 1p       & 1p+2p    \\
$\omega^+=-0.2$  & $0.231$  & $0.143$  & $0.203$   & $-0.203$  & $0.075$  & $0.115$  & $0.254$  & $0.221$  & $0.188$  & $0.253$  & $0.263$  \\
$\omega^+=-0.08$ & $0.324$  & $0.381$  &  $0.305$  & $0.319$   & $0.364$  & $0.375$  & $0.311$  & $0.312$  & $0.118$  & $0.312$  & $0.318$  \\
$\omega^+=0.11$  & $-0.097$ & $0.041$  & $-0.094$  & $0.146$   & $-0.022$ & $0.029$  & $-0.060$ & $-0.094$ & $0.033$  & $-0.081$ & $-0.066$  \\
\end{tabular}
\caption{$\DR$ for different harmonics of the Fourier series \eqref{eq:DTWfourier}, cases S3 and S6. Columns marked with 1SD (1OD) correspond to simulations where only the first same-direction (opposite-direction) wave is present, whereas 1p means the first harmonic pair combined, and 1p+2p the first two harmonic pairs combined.}
\label{tab:DRFourier}
\end{table}

\begin{table}
\centering
\begin{tabular}{l|c|ccccc|ccccc}
Case              & SIN     & S3      & 1SD      & 1OD      & 1p        & 1p+2p     & S6       & 1SD       & 1OD      & 1p       & 1p+2p   \\
$\omega^+=-0.2$   & $2.22$  & $2.09$  & $1.52$   & $0.33$   & $1.86$    & $2.00$    & $2.14$   & $2.02$    & $0.05$   & $2.07$   & $2.13$  \\
$\omega^+=-0.08$  & $1.46$  & $1.29$  & $1.00$   & $0.12$   & $1.12$    & $1.21$    & $1.42$   & $1.33$    & $0.04$   & $1.37$   & $1.41$  \\
$\omega^+=0.11$   & $1.42$  & $1.46$  & $0.97$   & $0.31$   & $1.29$    & $1.38$    & $1.43$   & $1.30$    & $0.08$   & $1.38$   & $1.42$  \\
\end{tabular}
\caption{$\Pin$ for different harmonics of the Fourier series \eqref{eq:DTWfourier}, cases S3 and S6. Columns as in Table \ref{tab:DRFourier}.}
\label{tab:PinFourier}
\end{table}

It remains to be established to what extent the (linear) superposition of the DTW harmonics can be useful in understanding the pattern of drag reduction and the differences between SIN and S3. 
To this aim, additional simulations (for cases S3 and S6) are run at three frequencies: $\omega^+=0.11$ (drag increase), $\omega^+=-0.08$ (drag reduction, where S3 performs better than SIN and S6 equals SIN), and $\omega^+=-0.2$ (drag reduction, where S3 performs worse than SIN, and S6 performs better). 
Instead of the actual DTW given by Eq.\ref{eq:DTW}, the employed forcing is sinusoidal, and contains one or more of the harmonics of the series \ref{eq:DTWfourier}. In particular, we consider the first two pairs of harmonics, isolated or in combination, for a total of eight additional simulations for each case. 

Hence, in these numerical experiments, the wall forcing is given by various combinations of the following sinusoids:
\begin{equation}
W(x,t;3)=\frac{3\sqrt{3}}{2\pi}A\Bigl[\sin(\omega t - \kappa_x x)+\frac{1}{2}\sin(\omega t + 2\kappa_x x) -\frac{1}{4}\sin(\omega t - 4\kappa_x x)  -\frac{1}{5}\sin(\omega t + 5\kappa_x x)\Bigr]
\end{equation}
for case S3, and
\begin{equation}
W(x,t;6)=\frac{3}{\pi}A\Bigl[\sin(\omega t - \kappa_x x)+\frac{1}{5}\sin(\omega t + 5\kappa_x x) -\frac{1}{7}\sin(\omega t - 7\kappa_x x)  -\frac{1}{11}\sin(\omega t + 11\kappa_x x)\Bigr]
\end{equation}
for case S6. 

Results from these experiments are shown in tables \ref{tab:DRFourier} and \ref{tab:PinFourier}, in terms of $\DR$ and $\Pin$ respectively. 
By focusing first on $\DR$, the first and relatively trivial observation is that the series \eqref{eq:DTWfourier} converges rather quickly in terms of drag reduction: for example at $\omega^+=-0.2$, the first pair of harmonics of S3 yields a drag reduction of 7.5\%, adding the second pair yields 11.5\% and the actual DTW at $m \to \infty$ yields 14.3\%.
A less obvious observation is the very different role played by the two harmonics at a given $m$, and their highly nonlinear combination. 
As an example, for the drag-reducing case at $\omega^+=-0.2$, the wave 1SD alone yields a significant 20.3\% drag reduction, whereas 1OD produces precisely the same amount of drag increase. However, the non-linear interaction between the two is such that, when the two harmonics are at play together, drag reduction prevails over drag increase with an outcome of $\DR=0.075$.  

\begin{figure}
\centering
\includegraphics[width=0.7\textwidth]{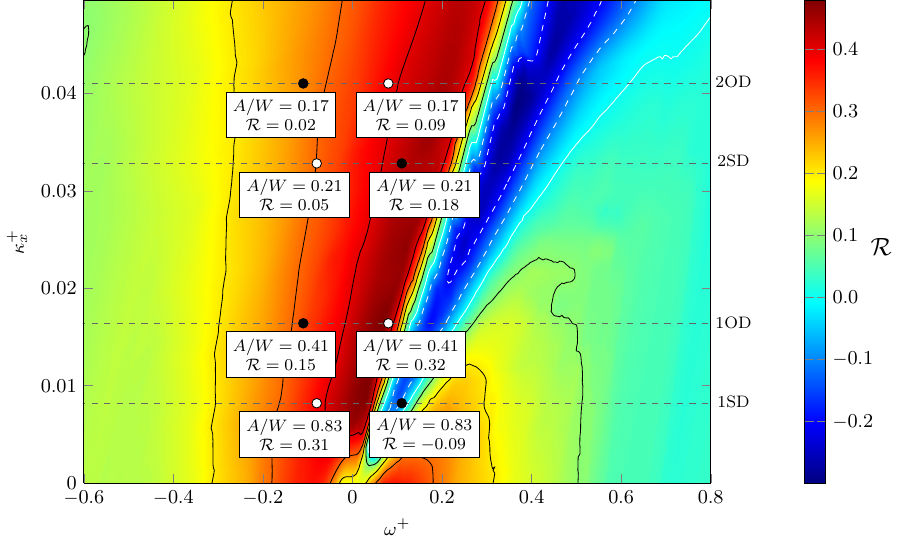}
\caption{Position on the drag reduction map of the first two harmonic pairs of the DTW with $s=3$, for a drag-reducing case (white dots: $\omega^+=-0.08$, $\kappa_x^+=0.0082$ and $\DR=0.38$) and a nearly drag-increasing one (black dots: $\omega^+=0.11$, $\kappa_x^+=0.0082$ and $\DR=0.04$). Note how each pair of corresponding harmonics of the same case (e.g. white dots for $\omega^+=0.08$) have the same frequency with opposite sign.
The drag reduction map is adapted from \cite{gatti-quadrio-2016}, for a plane channel flow at $Re_\tau=200$ and $A^+=14.2$. The legend below each dot quantifies the amplitude $A/W$ of the harmonic, and the drag reduction achieved by that harmonic alone.}
\label{fig:sines-on-plane}
\end{figure}

Considering just the first pair of harmonics explains some of the features observed in figure \ref{fig:RDisc}.
The amplitude of the 1SD component is closer to $A$ for S6 than S3, and the amplitude of 1OD decreases more rapidly for S6 than S3; the opposite happens for the wavenumber. By noting that, apart from the amplitude and a phase-shift, the 1SD wave for S6 correspond to the 2OD wave for S3, one understands why the S3 data are further away from SIN than S6 data, show stronger wiggles, and never achieve relaminarization.
For the two cases corresponding to the second and third row of table \ref{tab:DRFourier} (i.e. drag reduction and drag increase), the position of the first S3 harmonics in the plane of the control parameters is shown in figure \ref{fig:sines-on-plane}. The drag reduction map is recomputed from the information provided by \cite{gatti-quadrio-2016}, hence at $Re_{\tau}=200$ and for the planar geometry, and is adapted to the forcing intensity of the present case to convey a qualitative information of the behavior expected for the harmonic.
Results at $\omega^+=0.11$ are interesting to understand the effects of the drag-increasing components. The dominant wave travels at the same phase speed as the nominal wave and produces the same drag-increasing effect, quantitatively not too far from the one by SIN. 
However, as clearly shown by figure \ref{fig:sines-on-plane}, higher harmonics possess a different phase speed and a different wavenumber: they are all located in drag-reducing areas of the $\omega^+ - \kappa_x^+$ plane.  Hence, they act to weaken the drag-increasing effect of the nominal wave, resulting in a small drag reduction. The different 1OD wave between cases S3 and S6 explains why S6 instead achieves drag increase.

The case at $\omega^+=-0.08$ demonstrates an interesting interplay between the waves of the first pair. They lie both in the drag-reducing regime, but their combination results in an enhancement of $\DR$ for S3, while for S6 the drag variation is essentially unchanged. Once again, this happens because 1OD sits in a more effective region of the parameter space (in this case, nearly at the optimal position).
Lastly, at $\omega^+=-0.2$, the 1OD waves have an opposite effect: drag increase for S3, and drag reduction for S6. This affects the combined wave and reduces $\DR$ for S3, but increases it for S6.

Table \ref{tab:PinFourier} reports $\Pin$ for the same cases considered in table \ref{tab:DRFourier}. 
Again, the variation of the DTW compared to SIN can be ascribed to the different harmonic components, which differ in their amplitude and their position in the $\omega^+ - \kappa_x^+$ plane. 
In terms of power, however, the various harmonics are almost perfectly additive, in sharp contrast with drag reduction data. This is reasonable, as the input power for the spanwise forcing has little to do with the superimposed turbulent flow, and can be described quite well by the laminar transverse flow alone \citep{quadrio-ricco-2011}.

\begin{figure}
\centering
\includegraphics[scale=0.75]{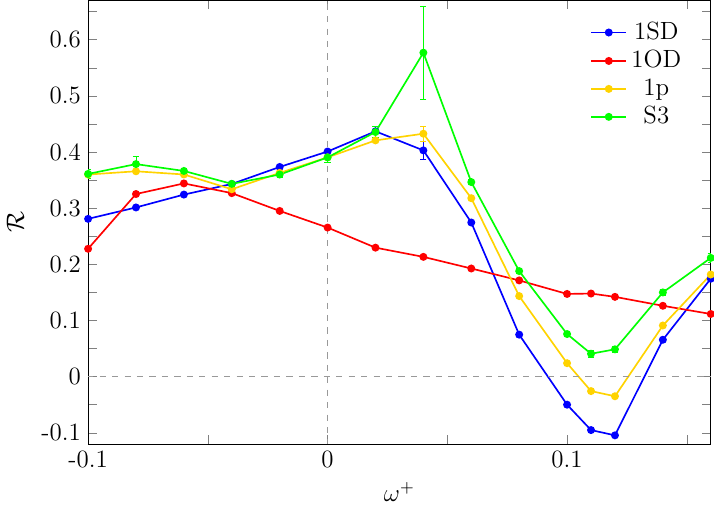}
\caption{Drag reduction rate $\DR$ computed with DNS for the first pair of harmonics (1SD and 1OD) and their combination (indicated with 1p) compared against S3.}
\label{fig:DRS3}
\end{figure}

A global view of the effect of the 1SD and 1OD waves, as well as their combination (indicated with 1p) is provided in figure \ref{fig:DRS3}, where results from an additional set of numerical experiments are plotted. For 15 points with frequency $-0.10 \le \omega^+ \le 0.16$, the drag reduction rate from case S3 is compared with the ones obtained with forcing by the sinusoidal wave 1SD, 1OD and their combination 1p, all with the amplitude prescribed by \eqref{eq:DTWfourier}.
First of all, one notices that the first pair of harmonics is responsible for most (but not all) of the effects of the DTW, especially at negative frequencies. The exception at $\omega^=+0.04$, where S3 nearly achieves laminarization but the 1p harmonics do not, has been already discussed.
The figure also provides detailed information on the non-trivial contribution of the isolated waves 1SD and 1OD. At first approximation, the 1SD curve resembles SIN, but for the reduced amplitude of the forcing; the 1OD curve, instead, acts at a doubled wavenumber, as previously exemplified in figure \ref{fig:sines-on-plane}, so that its horizontal axis is enlarged by a factor 2 and also reversed by virtue of the phase speed reversal.

Overall, the general picture emerging from this analysis is that the sinusoidal forcing is always best, as long as one can work with the optimal control parameters that yield maximum drag reduction. However, as soon as the forcing parameters do not assume their optimal values (something that is not inconceivable e.g. because of technological limitations), the higher harmonics of a non-sinusoidal forcing affect the outcome in a way that depend on their location in the parameter space, so that the discrete forcing may outperform the sinusoidal forcing. 
This is exactly the conclusion reached one decade ago by \cite{cimarelli-etal-2013}, who experimented with the temporal waveform for the spanwise oscillating wall. In both the spatial and temporal cases, regions of the parameters space exist where the sinusoidal forcing can be outperformed, in terms of both $\DR$ and $\NS$. However, such regions are far from the global optimum, in correspondence of which the sinusoidal forcing remains the best choice.

The final message is that, whenever the forcing is spatially distributed, its spatial discretization is an essential ingredient to evaluate drag reduction and its success in terms of energy savings. Experimental studies where a discrete form of spanwise forcing is employed can hardly be compared to results from a sinusoidal forcing, unless discretization is properly accounted for.

\subsection{Localized turbulence}
\label{sec:localized}

Once it is recognized that S3 and EXP data present the very same wiggles, which should not be attributed to experimental errors, but are a direct consequence of forcing discretization, one remaining difference becomes even more evident. 
In figure \ref{fig:RDisc}, EXP data do not reach above an apparent threshold of maximum $\DR \approx 0.33$, while S3, S6 and SIN achieve much higher values of $\DR$, up to the complete relaminarization of the flow.

\begin{figure}
\centering
\includegraphics[width=\textwidth]{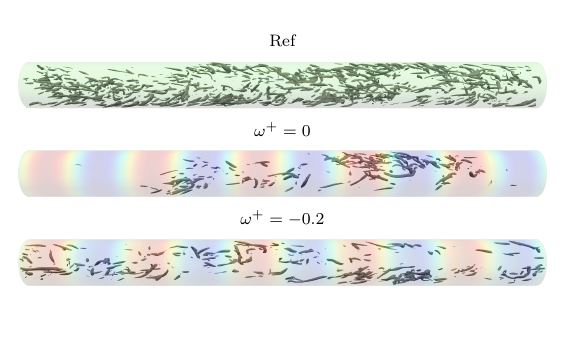}
\caption{Instantaneous snapshot of the reference and sinusoidally forced pipe flow, with parameters yielding drag reduction above ($\omega^+=0$) and below ($\omega^+=-0.2$) the apparent experimental threshold for $\DR$. The background colormap (from blu to red) encodes the azimuthal velocity of the wall; the grey isosurfaces are drawn for $\lambda_2^+ = -0.03$.}
\label{fig:snapshot}
\end{figure}

Even though the only experimental information provided by ABBCQ is the drag reduction value, the additional insight allowed by the present numerical study highlights a peculiar feature of the solution, which could explain the discrepancy. 
Instantaneous snapshots of the sinusoidally forced flow for two SIN cases with drag reduction above and below the apparent experimental threshold (namely at $\omega^+=0$ and $\omega^+=-0.2$, with $\DR=0.42$ and $\DR=0.23$ respectively) are compared to a snapshot from the reference flow in figure \ref{fig:snapshot}. The background colormap shows the azimuthal velocity at the pipe wall; isosurfaces for $\lambda_2$ \citep{jeong-etal-1997} are plotted at the level $\lambda_2^+=-0.03$, and provide a qualitative idea of the turbulent structures at play within the flow.
As expected, the control action reduces the presence of structures at both frequencies, with a more evident effect when drag reduction is higher. 
However, the most striking feature of the plot is that turbulent activity for $\omega^+ = 0$, where drag reduction is larger and goes above the apparent experimental threshold, appears to be spatially intermittent, whereas in the reference flow and also for the lower drag reduction case, it assumes a more conventional, spatially uniform look.
Turbulent structures at $\omega^+ = 0$ appear clustered, and surrounded by a refractory region where turbulence is absent or nearly so.
One can envisage the arrangement of the flow structures in what is usually called a puff \citep{barkley-2016, avila-barkley-hof-2023}, which is typical of transitional pipe flow; a notable difference here is that the flow is turbulent, the Reynolds number is relatively high, and the structures also undergo a swirling motion due to the spanwise flow induced by the forcing. 

Hence, we link the state of turbulent localization observed in DNS-computed flow fields to the ability to reach $\DR$ levels above the plateau of approximately 30 -- 35\% found in the experiment. This suggestion is supported by the observation that the local friction oscillates around the laminar value far from the turbulent regions, whereas in correspondence to them it becomes comparable to the experimentally measured value.
Several reasons might have prevented the ABBCQ pipe flow from reaching such a higher drag reduction state. Examples include the periodic streamwise boundary conditions as opposed to a fixed level of disturbance entering the experimental pipe via the inlet, or the idealized setting of the simulation, which is free from environmental disturbances.

\section{Conclusions}
\label{sec:conclusions}

In this work, direct numerical simulations have been used to replicate the successful turbulent pipe flow experiment by \cite{auteri-etal-2010} for skin-friction drag reduction based on spanwise forcing.
The experiment was carried out at a rather low value of the Reynolds number, which is replicated precisely here. We are aware that this is a marginally low $Re$, and in fact several control configurations lead to total of partial relaminarization of the flow. However, the study is not designed to address the important question of the $Re$ dependence on drag reduction (in general, and with spanwise forcing in particular), for which the discrete nature of the forcing should not play any crucial role. Rather, the goal here is to exploit the detailed comparison between experimental and simulation data obtained under nominally identical conditions to understand how a spatially discrete implementation of streamwise-travelling waves of azimuthal wall velocity affects the outcome of the flow control technique. In fact, any experimental implementation of the forcing unavoidably differs from the ideal sinusoidal function typically considered in numerical studies; accounting for the effect of spatial waveform is required for a proper assessment of the forcing performance.

The DNS simulations have been carried out with an original code that uses Fourier discretization for the homogeneous directions and resorts to compact finite differences for the radial one. Such mixed discretization is particularly suited for the turbulent pipe flow, and provides an efficient strategy to solve the issue of excess azimuthal resolution near the pipe axis, by allowing the number of azimuthal spanwise modes to gradually decrease as the axis is approached. Thanks to the efficiency of the code in terms of CPU and memory requirements, most of the present study has been carried out on a single Xeon Phi processor.

The experimental conditions have been replicated by implementing the spanwise forcing as a piecewise-continuous, streamwise- and time-periodic function. Differences between the drag reduction rate $\DR$ computed by DNS and measured in the experiment have been identified and explained.
The much higher drag reduction measured in simulations for both continuous and discrete forcing is suggested to derive from low-$Re$ effects, which enable the numerically simulated flow to reach a state of spatially localized turbulence and eventually fully relaminarize.
The apparently irregular behaviour of experimental $\DR$ is observed in the numerical results too, and is shown to derive from the discretization of the sinusoidal waveform, locally resulting into either larger of smaller $\DR$ depending on the combination of control parameters.  
By expanding the piecewise-continuous function into a Fourier series, the discrete forcing is shown to be equivalent to two families of sinusoidal waves, each with decreasing amplitude and wavelength as the order of the harmonics is increased: one family is made of waves travelling in the same direction of the discrete wave, whereas the other contains waves travelling in the opposite direction. Additional simulations have been run to observe the role of the various harmonics, concluding that discretization has a predictable but non-trivial effect depending not only on the degree of discretization, but also on the position of the nominal (sinusoidal) wave in the plane of the control parameters. In the end, the drag reduction performance of the discrete wave can be predicted from its harmonic content, provided full information for the sinusoidal waves is available 

To progress from the idealized setting of a DNS towards experimental or real-world applications, where a discrete spatial waveform is unavoidable, it is essential to fully understand the differences between continuous and discrete waveforms, and to exercise care when comparing DNS data computed for sinusoidal forcing with experimental, non-sinusoidal ones. 
This applies not only to raw drag reduction data, but also to the energetic requirements of the forcing.
Provided a comparison is carried out properly, a sinusoidal forcing remains the best option in correspondence of the optimal forcing parameters. Yet, whenever e.g. technological limitations prevent the actuator to operate in correspondence of the best forcing conditions, the discrete forcing can in principle outperform the sinusoidal one.

\section*{Declaration of Interests} 
The authors report no conflict of interest.

\section*{Acknowledgments}
The authors gratefully acknowledge the preliminary work carried out by Martina Biggi during her Master thesis work. 

\bibliographystyle{jfm}
	
\end{document}